\documentclass[a4paper]{article}
\usepackage{INTERSPEECH2019}
\usepackage[utf8]{inputenc}
\newcommand{\citep}{\cite}
\usepackage[hidelinks]{hyperref}
\usepackage{booktabs}
\usepackage{xcolor}
\usepackage{float}
\usepackage{siunitx}
\usepackage{tabularx}
\usepackage{booktabs}
\usepackage{subcaption}
\usepackage[acronym,nomain,nonumberlist]{glossaries}
\newacronym{hmms}{HMMs}{hidden Markov models}
\newacronym{hmm}{HMM}{hidden Markov model}
\newacronym{dnn}{DNN}{deep neural network}
\newacronym{dnns}{DNNs}{deep neural networks}
\newacronym{pase}{PASE}{problem-agnostic speech encoder}
\newacronym{tts}{TTS}{text-to-speech}
\newacronym{rmse}{RMSE}{root-mean-square error}
\newacronym{mcd}{MCD}{mel cepstral distortion}

\definecolor{burntorange}{rgb}{0.8, 0.33, 0.0}

\usepackage{soul}
\def\arraystretch{1.2}

\title{Problem-Agnostic Speech Embeddings for \\Multi-Speaker Text-to-Speech with SampleRNN}
\name{David Álvarez, Santiago Pascual, Antonio Bonafonte$^{\ast}$\thanks{$^\ast$A.~Bonafonte is currently at Amazon Research, Cambridge, UK.}}
\address{Universitat Politècnica de Catalunya}
\email{davidalvarezdlt@gmail.com, santi.pascual@upc.edu}

\begin{document}
    \maketitle
    \begin{abstract}
        \Gls{tts} acoustic models map linguistic features into an acoustic representation out of which an audible waveform is generated. The latest and most natural \gls{tts} systems build a direct mapping between linguistic and waveform domains, like SampleRNN. This way, possible signal naturalness losses are avoided as intermediate acoustic representations are discarded. Another important dimension of study apart from naturalness is their adaptability to generate voice from new speakers that were unseen during training. In this paper we first propose the use of problem-agnostic speech embeddings in a multi-speaker acoustic model for TTS based on SampleRNN. This way, we feed the acoustic model with speaker acoustically-dependent representations that enrich the waveform generation more than embeddings unrelated to these factors. Our first results suggest that the proposed embeddings lead to better quality voices than those obtained with one-hot embeddings. Furthermore, as we can use any speech segment as an encoded representation during inference, the model is capable to generalize to new speaker identities without retraining the network. We finally show that, with a small increase of speech duration in the embedding extractor, we dramatically reduce the spectral distortion to close the gap towards the target identities.
    \end{abstract}
    \section{Introduction}
        The fast development of new and deeper neural architectures has allowed the improvement of all tasks related to complex non-structured data. Speech synthesis is one of those tasks whose performance has improved dramatically with the latest trends in deep generative models. 
        \Gls{tts} is the technique by which a machine maps a text into audible speech. The task is especially difficult when we want to be able to generate speech from different speakers, forcing the system to learn how to differentiate among them but sharing some core information on the linguistic to acoustic signals correspondence. 
        
        Classic \Gls{tts} systems are based on statistical models that extract linguistic and prosodic features from text and map them into an acoustic representation to obtain the speech waveform~\citep{zen2007hmm, StatisticalParametric}. These models work in a two-stage methodology: (1) linguistic and prosodic features are used to predict the duration of phonetic units with a duration model (predicting number of acoustic frames to be predicted); and (2) these same linguistic, prosodic and duration features are then fed into an acoustic model to generate the actual acoustic units that will finally conform a waveform. These acoustic units usually take the form of spectral or cepstral features, and to make these statistical mappings in both stages we either rely on \gls{hmms} or \gls{dnns}~\citep{StatisticalParametric, ze2013statistical}. The latter outperform previous statistical systems, and as such most successful statistical models apply some type of neural component in both stages of the \Gls{tts}~\cite{zen2015unidirectional, pascual2016deep}. Remarkably, different neural architectures and optimization schemes can be applied to increase the generation efficiency and voice naturalness~\cite{pascual2018self, wu2015deep, valentini2015towards, wu2016investigating, uria2015modelling, zen2014deep}, especially to palliate over-smoothing effects that produce “buzziness” and muffled sounds ~\cite{StatisticalParametric}.
        
        The introduction of the WaveNet and SampleRNN systems showed that working on acoustic models at waveform level is plausible and currently outperforms acoustic feature-based systems in terms of naturalness~\citep{Wavenet,SampleRNN}. WaveNet was first used as a plain acoustic mapping conditioned on linguistic and prosodic features, whereas SampleRNN was first used as a vocoder in the Char2wav~\cite{Char2Wav} \Gls{tts} system. Nonetheless, both systems are flexible enough to be conditioned on many different features that either express linguistic contents or spectral features out of which waveforms are generated, as WaveNet showed in the Tacotron~\cite{Tacotron}~\Gls{tts}. Also SampleRNN has been used as an acoustic model~\cite{SampleRNNVoiceConversion} for voice conversion with linguistic conditionings.
        
        An important feature of statistical~\Gls{tts} systems, especially related to our work, is their flexibility in terms of modeling different speaker characteristics to mimic them in the generation. Acoustic models and neural vocoders have the capacity to model different speaker identities within the same neural structure. If we model all of them in the same training session, the way in which these multiple identities can be controlled is with a specific multi-output structure, where each output generates the acoustics of its respective speaker~\cite{SantiMultiSpeaker, DnnTtsMulti}, or with some input codes or embeddings expressing the modeled identity in a single model output~\cite{Wavenet, gibiansky2017deep}. In case of having new identities to model after the model is already trained, there exist a number of speaker adaptation techniques to tune models to new identities with the least amount of data possible. These techniques usually rely on some form of model fine-tuning~\cite{wu2015study, SantiMultiSpeaker, DnnTtsMulti}, but adapting identities with tiny amounts of data should potentially rely on more descriptive speaker embeddings than one-hot ones. This way, we could extract the embeddings from an example utterance of a target identity, and we could avoid fine-tuning the models by injecting the extracted acoustic descriptors of the target speaker~\cite{DBLP:journals/corr/abs-1809-10460, TransferLearningSpeaker}. Even though this is a desirable case, the best results are still obtained after a fine-tuning stage of the model parameters even in data scarcity situations.
        
        In this paper, we first build a multi-speaker acoustic model with a SampleRNN conditioned on linguistic-prosodic features and speaker identity embeddings. We then propose the use of problem-agnostic acoustic embeddings extracted from chunks of speech that serve as seed examples of speaker identity traits. This brings us the possibility of generating new identities by obtaining the acoustic vectors for new speakers without fine-tuning the network. Finally, we present an initial study of speaker adaptation without fine-tuning by varying the amount of available seed speech for different test speakers that remain unseen during training. 
        
        This work is structured as follows: we first review SampleRNN in section~\ref{sec:samplernn} as our waveform generator with which we build the acoustic model. In section~\ref{sec:pase} we review the \gls{pase}, with which we obtain the speaker identity embeddings. The experimental setup is described in section~\ref{sec:exp_setup}, where datasets, modeling features and hyper-parameters are described. Results and conclusions are discussed in sections~\ref{sec:results} and~\ref{sec:conclusions}, respectively. The code of this project is publicly available online~\footnote{\url{https://github.com/davidalvarezdlt/samplernn_pase}}.
       
    \section{SampleRNN for Text-to-Speech}
        \label{sec:samplernn}
\begin{figure}[t]
            \centering
            \includegraphics[width=0.25\textwidth]{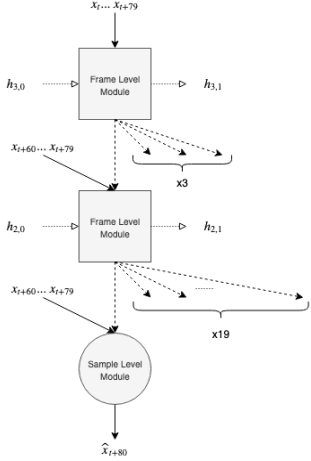}
            \caption{Original SampleRNN architecture with $2$ frame-level layers and upsampling ratios \{4, 20\}}
            \label{fig:samplernn}
        \end{figure}

\begin{figure}[t]
            \centering
            \includegraphics[width=0.45\textwidth]{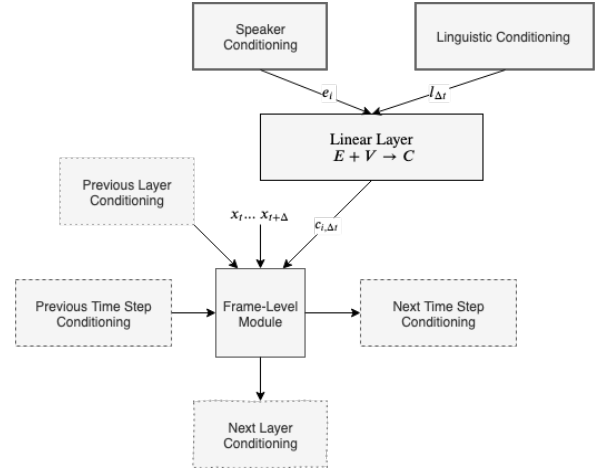}
            \caption{Combination of speaker and linguistic features along with other SampleRNN model-specific conditioning inputs}
            \label{fig:samplernn_conds}
        \end{figure}

        The original design of SampleRNN \cite{SampleRNN} is an auto-regressive model with data that propagates both from upper to bottom layers and from previous to next time steps (see Fig.~\ref{fig:samplernn}). In each layer, it uses different scales of previous samples to condition the lower layer until it predicts the sample itself in the last layer. In the original work, the lowest layer is called sample-level layer and the others are called frame-level layers.
        As we want to control both the content of the speech and the speaker identity, we must somehow inject conditioning information to the system in addition to previous samples. 
        We first experimented injecting the conditionings at the top-level layer (as in existing vocoding applications of SampleRNN~\citep{MultiSpeakerNeuralVocoder}), however this solution had difficulties to generate intelligible speech in our acoustic model for multiple identities.
        To solve that, we use a similar strategy as in~\citep{SampleRNNVoiceConversion}, combining both the original inputs with a global feature vector composed of speaker and linguistic features. These new feature vectors are used as inputs in the frame and sample level layers and are then concatenated with the other inputs before getting into the recurrent units.
        Hence, given a speaker $i$ with its feature vector $\pmb{e}_i \in \mathbb{R}^{E}$ and the linguistic feature vector of a certain time interval $\pmb{l}_{\Delta t} \in \mathbb{R}^{V}$, we use a linear layer to obtain a global conditioning vector $\pmb{c}_{i,\Delta t} \in \mathbb{R}^{C}$. Fig.~\ref{fig:samplernn_conds} gives an overall scheme of this conditioning methodology along with other inputs of frame-level layers.
        The initial states of the frame-level layers $h_{z,0}$ are learned via back-propagation. For the case of the sample-level module, it follows a similar structure but without time-step conditioning.
        Speaker identity can be introduced either with embedded one-hot codes as in previous approaches~\cite{Wavenet}, or with some feature extractor out of speech signals that serve as seeds~\cite{DBLP:journals/corr/abs-1809-10460, TransferLearningSpeaker}. In this work we propose to use a problem-agnostic speech encoder to follow the latter scheme, as described next.
    
    \section{Problem-Agnostic Acoustic Embeddings}  
        \label{sec:pase}
        
        As mentioned earlier, the identity conditionings injected to SampleRNN can be one-hot codes, as typical; or acoustically relevant features extracted from some spoken examples from the modeled identities that serve as seeds. 
        
        To extract these acoustic features, we use the problem-agnostic speech encoder (PASE)~\cite{PASE}. PASE is an encoder that transforms raw waveforms $\pmb{x}$ into an intermediate representation $\pmb{z}$ with a deep fully convolutional structure. This converts a waveform into a collection of 100-dimensional frames with frame rate 100Hz. The $\pmb{z}$ representation is learned out of a collection of self-supervised tasks that predict different factors of the input waveform: the waveform itself, its log-power spectrum, the MFCC coefficients, etc. Each prediction is perform by a small multi-layer perceptron (MLP), called worker. 
        
        We decide to use PASE for its proven ability to capture identity and prosodic cues, as shown in the original paper with the results on speaker identification and emotion recognition~\cite{PASE}. In our work, we use the publicly available pre-trained encoder to extract our acoustic seeds, which are replaced by the usual one-hot embeddings.
        
        Let the function $\Phi(\pmb{x})$ represent the output latent representation $\pmb{z}$ of the PASE encoder. We average this output vectors in time to obtain a summary of the speaker characteristics for a certain seed signal length $T$, thus obtaining the embeddings $\boldsymbol{e}_i$ as:
        
        
        \begin{equation}
            \boldsymbol{e}_i = \frac{1}{L}\sum_{n=0}^{L-1} \Phi(\boldsymbol{s}_{i})[n]
        \end{equation}
        Where $\boldsymbol{s}_{i}$ corresponds to the $i$-th speaker seed signal, and $L$ is the number of PASE frames coming out of the encoding (a 16\,kHz signal the waveform is decimated to a 100\,Hz signal after $\Phi(\boldsymbol{s}_{i})$).

    \section{Experimental Setup}
        \label{sec:exp_setup}
        \subsection{Datasets}
            \label{sec:datasets}

            For our experiments, we use utterances from two different datasets: VCTK~\cite{VeauxVCTK} and CMU Arctic~\cite{kominek2003cmu}. We decide to use speakers from both datasets as, while CMU Arctic contains more recorded speech per speaker, it does not contain enough speakers for our experiments. To avoid excessively modelling of silences, we trim them to a maximum length of 100\,ms with the help of a voice activity detector.
            
            

            To train the base models, we select the 20 speakers of each gender with most speech duration, allocating a total of 40 speakers for this purpose. Then, for each of those speakers, we select a total of 45\,s of speech for the validation split and another 45\,s to perform objective tests. The remaining data of each speaker is then allocated to the training split, obtaining an unbalanced training dataset where the less representative speaker (VCTK speaker) has about 10 minutes of speech and the most representative one (CMU Arctic speaker) reaches the 64\,min. In total, the sum of the individual contributions of the 40 speakers reaches the 12\,h of training data and 30\,min. of validation and test data.
            
            For the adaptation phase, we select 5 different random speakers of each gender (a total of 10 speakers) and allocate enough data in the training split to experiment with seed signal lengths $T$ up to 120\,s. Note that we do not adjust the model weights, but we still require a training set of utterances to have enough $T$ samples to build the acoustic seed. We then assess the adapted speaker similarity towards the target identity using a different test split with 180\,s of data per speaker and the metrics described in section~\ref{sec:objective}.
        
        \subsection{Linguistic-prosodic features}
            As mentioned in section~\ref{sec:samplernn}, one of the conditioning features that we input in every frame and sample level module is related to the linguistic and prosodic contents, as we are building an acoustic model. In our experiments, we will compare the performance difference when using two different sets of features:
            \begin{itemize}
                \item Linguistic features with a forced duration model.
                \item The previous linguistic features with $\log F0$ and unvoiced/voiced (UV) binary flags (similar to~\cite{Wavenet, SampleRNNVoiceConversion}).
            \end{itemize}
            Given the set of utterances of a certain speaker, we use Merlin~\cite{Merlin} to extract a total of $53$ linguistic-prosodic features per phonetic unit jointly with its forced duration. Then, we repeat these features in time to fit the waveform time resolution $\Delta t$ using the time-stamps as reference to include the duration information to the feature vector:
            \begin{itemize}
                \item Absolute duration: the total duration of the phoneme.
                \item Relative duration: the position of the generated $\Delta t$ inside the phoneme, ranging from 0 to 1. 
            \end{itemize}
            We use the Ahocoder vocoder to extract $\log{F0}$ and V/UV features~\cite{erro2013harmonics}. All real-valued features (either linguistic or prosodic) are z-normalized with speaker-dependent statistics except for the relative duration, which is already bounded.
        
        \subsection{Experiments} 
            \label{sec:experiments}
            First, we want to assess the performance difference between using the PASE embeddings against plain one-hot embeddings. Secondly, we want to make a first evaluation of the adaptation capability of the PASE embeddings by picking different seed signal lengths $T$ to infer the identity embedding $\boldsymbol{e}_i$. 
            
            For the first experiment, four SampleRNN variations are trained for $50$ epochs,
            as the result of using $i)$ one-hot vs. PASE embeddings and $ii)$ including or not $\log F0$ + UV
            
            The training PASE embeddings are extracted from $T=60\,s$ of seed speech per speaker. Then, we analyze the performance of each model in terms of the negative log-likelihood loss and finally extract the objective distortion metrics of section~\ref{sec:objective}. 
            
            On the other hand, to assess the speaker adaptation, we pick the 10 randomly sampled speakers described in section~\ref{sec:datasets} and the PASE embedding experiment without $\log{F0}$ information from the previous list. For each new speaker, we randomly sample chunks of lengths $T = \{1, 10, 60, 120\}$ seconds to build their time-dependent acoustic PASE embedding. Then, we compute the objective distortion of these speakers' synthesized test utterances for each one of the 4 embeddings against their own ground-truth utterances.
        \subsection{Hyperparameter tuning}
            The hyper-parameter configuration used throughout all the experiments is summarized in Table~\ref{tab:hyperparameters}. Note that, depending on the experiment, only the PASE seed or the speaker embedding is used to represent the speaker identity. The learning rate is reduced on validation plateaus. The optimizer is Adam with all default parameters except for the learning rate. To have a reference on the computational cost per experiment, each model is ran on a Titan X GPU for approximately 8 days to reach the 50 epochs.
            \begin{table}[H]
                \renewcommand{\arraystretch}{1}
                \caption{Hyperparameters used in the experiments}
                \begin{tabular}{lc}
                    \toprule
                    {Speech sampling frequency} & $16\ \si{\kilo\hertz}$ \\
                    {Speech quantization} & 8 bits ($\mu\text{Law}$)  \\ 
                    {Speaker embedding size} ($E$) & $100$  \\
                    {Global features size} ($C$) & $50$ \\
                    {Categorical linguistic features embedding size} & $15$ \\
                    {Top frame-level seq. length} & $13$ \\
                    {Top frame-level inp. size} ($\Delta t$) & $80$ \\
                    {Upsampling Ratios} & $4, 20$ \\
                    {GRU hidden size} & $1024$ \\
                    {Batch size} & $128$ \\
                    {Initial learning rate} & $10^{-4}$ \\
                    {Learning rate patience} & $3$ \\
                    {Learning scaling factor} & $0.5$ \\\bottomrule
                \end{tabular}
                \label{tab:hyperparameters}
            \end{table}
            \begin{figure*}[htpb!]
                \centering
                \begin{subfigure}{.65\columnwidth}
                    \includegraphics[width=\columnwidth]{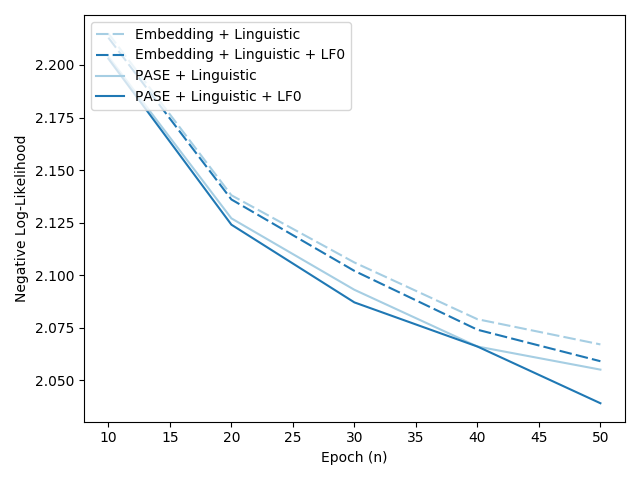}
                    \caption{Validation Split - Loss}
                    \label{fig:results_loss}
                \end{subfigure}
                \hfill%
                \begin{subfigure}{.65\columnwidth}
                    \includegraphics[width=\columnwidth]{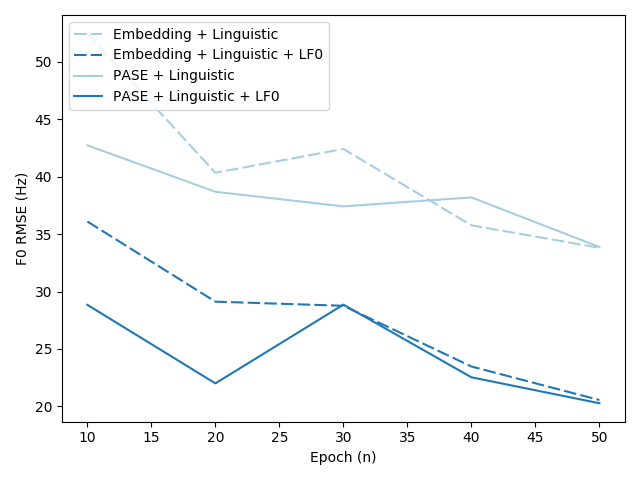}
                    \caption{Test Split - RMSE of F0}
                    \label{fig:results_f0}
                \end{subfigure}
                \hfill%
                \begin{subfigure}{.65\columnwidth}
                    \includegraphics[width=\columnwidth]{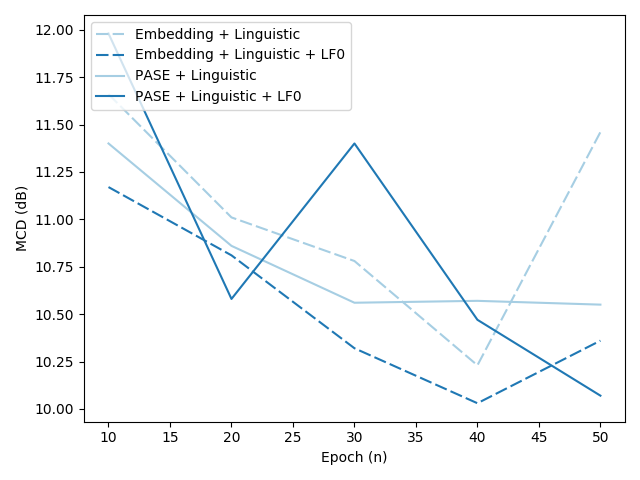}
                    \caption{Test Split - MCD}
                    \label{fig:results_mcd}
                \end{subfigure}
                \caption{Objective measures of the acoustic multi-speaker TTS model}
                \label{fig:results_acoustic}
            \end{figure*}
        \subsection{Objective Metrics}
            \label{sec:objective}
            We objectively evaluate the performance of our experiments by comparing the original utterances with the ones generated by our systems in two terms. First we have direct likelihood metrics as SampleRNN is a classifier computing the probability of the next waveform sample given the previous ones, which means that we get a score correlated to the generated quality in the likelihood validation and test curves. Secondly, we have spectral distortion measures that depict how good is the waveform generation embedding speaker, prosodic and content characteristics in the speech (without an explicit modelling of spectral features themselves). Hence, we measure the \gls{mcd} (in dB) and the RMSE of F0 (in Hz), as usually done in acoustic models assessment for \gls{tts}~\cite{kubichek1993mel, pascual2016deep}.
    \section{Results}
        \label{sec:results}
        Fig.~\ref{fig:results_acoustic} illustrates the likelihood convergence in validation and the acoustic distortion metrics in test for the four different \gls{tts} variations explained in section~\ref{sec:experiments}. We chose to train each model for 50 epochs (as stated in section~\ref{sec:experiments}) for comparability purposes, but the different models can converge further as shown in these results. Nonetheless, some evidences appear at this point worth commenting. First, we can clearly see that the two experiments using our acoustic embeddings converge quicker than with the standard approach overall both in terms of validation likelihood and acoustic distortions. The peaks we observe at some points in the spectral distortions are probably a reflect of under-optimized models, but we can still observe a clear convergence acceleration trend in the validation loss for models involving PASE embeddings that make these \gls{tts} variations comparable.
        
        Moreover, the distortions of the models with $\log F0$ contours as inputs are lower than those that lack them. This is expected because these contours supply long-term information that is useful for SampleRNN to retain far-past information better than it might in its internal states, as observed first in~\cite{Wavenet}. Nevertheless, the experiment with PASE embeddings that lacks $\log F0$ generally obtains better results than those of the one-hot embeddings with the $\log F0$ contours. We hypothesize that this happens due to PASE embeddings capturing some bias in each speaker prosodic traits, but we still have an averaged representation in this case so it does perform worse than when we use PASE embeddings with the prosodic contours where dynamic long-term information is still fully present.
        
\begin{figure}[t]
            \centering
            \includegraphics[width=0.40\textwidth]{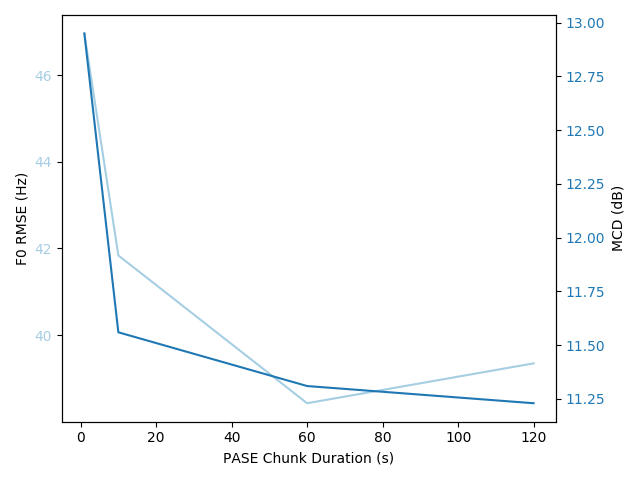}
            \caption{MCD and RMSE of $F0$ for new speakers with respect to the length of speech used for the embedding generation}
            \label{fig:results_pase}
        \end{figure}

        For the sake of simplicity, in our first adaptability experiment we only take the PASE embeddings (without $\log F0$ contours) model to proceed to the next experiment. After inferring four embeddings per speaker identity (one per seed length as introduced in section~\ref{sec:experiments}) we use them to synthesize the test utterances. Then, the 10 different speaker distortion curves for both RMSE and \gls{mcd} are averaged and depicted in Fig.~\ref{fig:results_pase}. There we observe that the more seed signal we have, the less distortion we obtain across speakers, especially for \gls{mcd}. The decreasing distortion trend depending on the amount of seed speech shows so far a promising direction towards generalizing speaker identities without fine-tuning \gls{tts} models. However, better optimized models are a potential need as shown in these results as models could converge further, hence biasing the adaptation curves to lower distortion rates with the same seed lengths. Qualitative results are available online in an audio samples webpage~\footnote{\url{http://veu.talp.cat/samplernn_pase}}. Informal listenings indicate that PASE embedding improve slightly the quality and that systems are significantly better if F0 is included as input. The system proposed allows to generate speech with the identity of unseen speakers providing just the PASE embedding. As a general rule, 10\,s are required to capture the identity. On the other hand, for the quality, it benefits from embeddings computed using longer speech segments (10-60\,s).  
            
    \section{Conclusions}
    \label{sec:conclusions}
        In this work we have proposed the use of problem-agnostic speech embeddings as an acoustically meaningful representation of speaker identities for a multi-speaker SampleRNN based acoustic model for \gls{tts}. We have concluded that the use of these new embeddings help the network to converge quicker and obtain better objective results in terms of likelihood and spectral distortions.
        Additionally, we also perform a first step towards speaker adaptation by inferring new identity embeddings for speakers unseen during training from some seed speech without any fine-tuning involved.
        Future lines of research further developing the contents presented here include training the models for longer periods and to fine-tune the \gls{pase} encoder simultaneously to the \gls{tts} training such that VCTK and CMU recording conditions can be captured along with the \gls{tts} learning.

    \section{Acknowledgements}
        This research was supported by the project TEC2015-69266-P (MINECO/FEDER, UE).
        
    
    \bibliography{refs}
\end{document}